\documentclass[12pt]{article}
\usepackage{amsfonts,amssymb,amsmath}
\usepackage{epsfig}
\usepackage[all]{xy}

\usepackage[usenames,dvipsnames]{color}
\usepackage{enumerate} 
\usepackage{ulem}

\newcommand\RR{\mathbb R}
\newcommand\CC{\mathbb C}

\newcommand{\Mt}{{\mathcal M}}

\newcommand\beq{\begin{equation}}
\newcommand\eeq{\end{equation}}

\newtheorem{theorem}{Theorem}
\newtheorem{lemma}{Lemma}

\newtheorem{remark}{Remark}
\newtheorem{proposition}{Proposition}

\begin{document}

\title{{Moutard transforms for the conductivity equation}
\thanks{The main part of the work was fulfilled during the visit of the 
first author to the Institut des Hautes \'Etudes Scientifiques
in November 2017 and to the Centre de Math\'ematiques Appliqu\'ees of \'Ecole Polytechnique in September-October 2018. The first author was partially supported by the Russian Foundation for Basic Research, grant 17-51-150001 ``Quasilinear equations, inverse problems, and their applications''. The second author was partially supported by PRC 1545 CNRS/RFBR: ``\'Equations quasi-lin\'eaires, probl\`emes inverses et leurs applications''.}}

\author{P.G. Grinevich
\thanks{L.D. Landau Institute for Theoretical Physics,
pr. Akademika Semenova 1a, 
Chernogolovka, 142432, Russia; Lomonosov Moscow State University, Faculty of Mechanics and Mathematics, 
Russia, 119991, Moscow, GSP-1, Leninskiye Gory 1, Main Building;
e-mail: pgg@landau.ac.ru} \and R.G. Novikov\thanks
{CNRS (UMR 7641), Centre de Math\'ematiques Appliqu\'ees, 
\'Ecole Polytechnique, 91128, Palaiseau, France; IEPT RAS, 117997, Moscow, Russia;
e-mail: novikov@cmap.polytechnique.fr}}
\date{}
\maketitle
\begin{abstract}
We construct Darboux-Moutard type transforms for the two-dimensional conductivity equation. This result continues our recent studies of 
Darboux-Moutard type transforms for generalized analytic functions. In addition, at least, some of the Darboux-Moutard type transforms of the 
present work admit direct extension to the conductivity equation in multidimensions. Relations to the Schr\"odinger equation at zero energy are also shown.
\end{abstract}



\section{Introduction}

We consider the two-dimensional isotropic conductivity equation:
\begin{equation}
\label{eq:hc1}
\mbox{div}\big({\sigma(x)}\nabla u(x) \big)=0, \ \ x=(x_1,x_2), \ \ x\in D\subseteq\RR^2, 
\end{equation}
where $D$ is an open bounded domain in $\RR^2$.
This equation arises in different physical context; see, for example,  \cite{LL7}, \cite{LL8}. In particular, in electrical 
problems $\sigma(x)$ is the elecrical conductivity in $D$, and  $u(x)$ is the electric potential. In solid state 
thermal problems $\sigma(x)$ is the heat conductivity, and $u(x)$ is the temperature.

In the present work we show that the conductivity equation~(\ref{eq:hc1}) admits Moutard-type transforms, going back to \cite{Mout}. 
Such transforms were successfully used in studies of integrable systems of mathematical physics and differential geometry, in spectral 
theory and in complex analysis; see, for example, \cite{MatvSal}, \cite{NS}, \cite{YLW}, \cite{TT}, \cite{NTT}, \cite{Taim1}, \cite{Taim2}, \cite{GRN1},  \cite{GRN2},  \cite{GRN3}, \cite{MT}, \cite{NT}, \cite{NT2}. In particular, the present article can be considered as a direct continuation of our recent 
works \cite{GRN1}-\cite{GRN3} on Moutard-type transforms for the generalized analytic functions. In turn, works \cite{GRN1}-\cite{GRN3} were 
stimulated by \cite{Taim1}, \cite{Taim2}. 

We recall also that equation (\ref{eq:hc1}) is the continuity equation
\begin{equation}
\label{eq:conj1}
\partial_{x_1} I_1 + \partial_{x_2} I_2 =0,\ \ x\in D,
\end{equation}
for the current density $I$,
\begin{equation}
\label{eq:hc1.1}
I(x) = {\sigma(x)}\nabla u(x), \ \ I=(I_1,I_2), \ \ x\in D.
\end{equation}
Assuming that $D$ is simply connected, we consider also the stream function $v$ for $I$, where
\begin{equation}
\label{eq:conj1.2}
\partial_{x_2} v = I_1, \ \  -\partial_{x_1} v=I_2,\ \ x\in D,
\end{equation}
(see, for example, \cite{LL6}), where integration constant may depend on the particular situation. 

The stream function $v$ satisfies the following equation:
\begin{equation}
\label{eq:conj1.3}
\mbox{div}\big({\sigma^{-1}(x)}\nabla v(x) \big)=0, \ \ x=(x_1,x_2), \ \ x\in D\subseteq\RR^2.
\end{equation}
In literature equation (\ref{eq:conj1.3}) is sometimes considered as the conjugate equation to (\ref{eq:hc1}); see, for example, \cite{CRW}.

In particular, in the present work we use the fact that equation~(\ref{eq:hc1}) can be written as a reduction of the following two-dimensional Dirac equation (see \cite{BU}):
\begin{equation}
\label{eq:hc2}
\left[
\left(\begin{array}{cc} \partial_{\bar z} &  0\\ 0  &  \partial_{z} \end{array}\right) -
\left(\begin{array}{cc}  0 &   q \\ \bar{q}  &  0 \end{array}\right)
\right]\left(\begin{array}{cc} \psi_1 \\ \psi_2  \end{array}\right) =0 \ \ \mbox{in} \ \ D,
\end{equation}
where 
\begin{equation}
\label{eq:hc3}
\begin{split}
\partial_{z} = \frac{1}{2}\left(\partial_{x_1}-i\partial_{x_2}\right), \ \ 
\partial_{\bar z} = \frac{1}{2}\left(\partial_{x_1}+i\partial_{x_2}\right), \\
\ \ q=q(x), \ \ \psi_{j}=\psi_{j}(x), \ \ j=1,2, \ \  x=(x_1,x_2).
\end{split}
\end{equation}
We recall that if $u$ satisfies (\ref{eq:hc1}), then
\begin{equation}
\label{eq:hc4}
\psi_1 = \sigma^{1/2} \partial_{z} u, \ \ \psi_2 = \sigma^{1/2} \partial_{\bar z} u,
\end{equation}
satisfy  (\ref{eq:hc2}), where
\begin{equation}
\label{eq:hc5}
q = -\frac{1}{2}\partial_{z}\log(\sigma), \ \ \bar{q} = -\frac{1}{2}\partial_{\bar z}\log(\sigma).
\end{equation}
We use also that (\ref{eq:hc2}) is equivalent to the following equation:
\begin{equation}
\label{eq:hc6}
\partial_{\bar z} \psi = q \overline{\psi} \ \ \mbox{in} \ \ D,
\end{equation}
which is the basic equation of the generalized analytic functions theory (see \cite{Vek}). More precisely:
\begin{enumerate}[(i)]
\item if $\psi_1$, $\psi_2$ satisfy (\ref{eq:hc2}), then $\psi_+ =\frac{1}{2}(\psi_1+\overline{\psi_2})$ 
and $\psi_- =\frac{1}{2i}(\psi_1-\overline{\psi_2})$ solve  (\ref{eq:hc6});
\item if $\psi_+$,  $\psi_-$ satisfy (\ref{eq:hc6}), then $\psi_1=\psi_++i\psi_-$, $\psi_2=\overline{\psi_+}+i \overline{\psi_-} $ solve  (\ref{eq:hc2}).
\end{enumerate}

The property that $\psi_1$, $\psi_2$ and $q$ in  (\ref{eq:hc2}) admit representations (\ref{eq:hc4}), (\ref{eq:hc5}) implies a non-trivial reduction of
equation (\ref{eq:hc2}). The compatibility of this reduction with the Moutard-type transforms from \cite{GRN1}-\cite{GRN3} is established in the present 
article. 

The main results of the present work are given in Sections~\ref{sec:3} and \ref{sec:5}. In these Sections we construct and study  Moutard-type transforms for the two-dimensional conductivity equation (\ref{eq:hc1}). In addition, in Section~\ref{sec:5} we show that, at least, some of these Moutard-type transforms admit direct extension to the conductivity equation in multidimension.  Besides, in Section~\ref{seq:schr} we continue studies of the Moutard-type transforms constructed in Sections~\ref{sec:3} and \ref{sec:5} in the framework of relations between the multidimensional conductivity equation and the Schr\"odinger equation at zero energy. 

Note that, for the case of bounded domain $D$, possible natural analytical assumptions on $q(x)$, $\sigma(x)$ are as follows:
\begin{align}
&q \in L^p(D),\label{eq:assumpt1}\\
&\sigma \in W^{1,p}(D),\label{eq:assumpt2}\\
&0 < \sigma_0 \le \sigma(x) \le \sigma_1 < +\infty,\label{eq:assumpt3}
\end{align}
where $p>2$. Actually, assumption (\ref{eq:assumpt1}) is essential in the standard theory of generalized analytic functions, see \cite{Vek}, and assumption  (\ref{eq:assumpt3}) is essential in the standard mathematical theory of the conductivity equation for the ellipticity.

Note also that the Moutard-type transforms constructed in the present work permit us, in particular, to study equation~(\ref{eq:hc1}) not only under assumptions (\ref{eq:assumpt2}), (\ref{eq:assumpt3}), but also for singular conductivities $\sigma$, not satisfying (\ref{eq:assumpt3}), in a way similar to the approach used in \cite{GRN1}-\cite{GRN3} to study generalized analytic functions with contour singularities. 

\section{Simple Moutard transforms for generalized analytic functions}
\label{sec:2}

Following \cite{Vek}, \cite{GRN1}-\cite{GRN3}, we consider the pair of conjugate equations of the generalized analytic function theory:  
\begin{align}
\label{eq:gan1}
&\partial_{\bar z} \psi = q \bar \psi \ \ \mbox{in} \ \ D,\\
\label{eq:gan2}
&\partial_{\bar z} \psi^+ = -\bar q \bar \psi^+ \ \ \mbox{in} \ \ D,
\end{align}
where $\partial_{z}$, $\partial_{\bar z}$ are defined in (\ref{eq:hc3}), $z = x_1 + i x_2$, $\bar z =x _1 - i x_2$,  $D$ is an open simply connected domain in 
$\CC\cong\RR^2$,  $q=q(z)$ is a given function in $D$. In addition, in this article the notation $f=f(x)=f(z)$ does not mean that $f(z)$ is holomorphic function 
in $z$ unless it is explicitly specified.

Next, as in \cite{Vek}, \cite{GRN1}-\cite{GRN3}, we associate with a pair of functions $\psi$, $\psi^+$, satisfying (\ref{eq:gan1}), (\ref{eq:gan2}), respectively, 
the following imaginary-valued potential $\omega_{\psi,\psi^+}$ defined by:
\begin{equation}
\label{eq:k1}
\partial_{z} \omega_{\psi,\psi^+} =\psi\psi^+, \ \ 
\partial_{\bar z} \omega_{\psi,\psi^+} =-\overline{\psi\psi^+} \ \ \mbox{in} \ \ D,
\end{equation}
where the pure imaginary integration constant may depend on the particular situation. We recall that the compatibility of (\ref{eq:k1}) 
follows from (\ref{eq:gan1}), (\ref{eq:gan2}). 

Let $f$, $f^+$ be some fixed solutions of equations (\ref{eq:gan1}), (\ref{eq:gan2}), respectively, with given $q$. Then
a simple Moutard-type transform $\mathcal{M}=\mathcal{M}_{q,f,f^+}$ for the pair of conjugate equations (\ref{eq:gan1}), (\ref{eq:gan2}) is given by the formulas
(see \cite{GRN1}-\cite{GRN3}):
\beq
\label{eq:m3}
\tilde q = \mathcal{M} q= q + \frac{f\overline{f^+}}{\omega_{f,f^+}},
\eeq
\beq
\label{eq:m1}
\tilde\psi=\mathcal{M} \psi=
\psi-\frac{\omega_{_{\psi,f^+}}}{\omega_{f,f^+}}\,f , \ \ 
\tilde{\psi}^+=  \mathcal{M} \psi^+=  \psi^+ - \frac{\omega_{f,\psi^+}}{\omega_{f,f^+}}\, f^+,
\eeq
where $\psi$, $\psi^+$ are arbitrary solutions of (\ref{eq:gan1}) and (\ref{eq:gan2}).

The point is that the functions $\tilde\psi$, $\tilde\psi^+$ defined in (\ref{eq:m1})
satisfy the conjugate pair of Moutard-transformed equations (see \cite{GRN1}-\cite{GRN3}):
\begin{align}
\label{eq:gan3}
&\partial_{\bar z} \tilde\psi= \tilde q\, \overline{\tilde\psi}& \ \ &\mbox{in} \ \ D, \\ 
\label{eq:gan4}
&\partial_{\bar z} \tilde\psi^+= -\overline{\tilde q}\, \overline{\tilde\psi^+}& \ \ 
&\mbox{in} \ \ D,
\end{align}
where $\tilde q$ is defined in (\ref{eq:m3}). In addition (in the simplest case), if $D$ is a simply connected open bounded domain with $C^{1}$-boundary, $q$ satisfies (\ref{eq:assumpt1}) and $f, f^+\in W^{1,p}(D)$, and $\omega_{f,f^+}\ne 0$ in $D\cup \partial D$, then the transformed coefficient $\tilde q$ satisfies (\ref{eq:assumpt1}) as well as the initial $q$ (see \cite{GRN1}). On the other hand, the Moutard-type transforms (\ref{eq:m3}), (\ref{eq:m1}) permit to create and remove contour singularities in $q, \psi, \psi^+$, see \cite{GRN1}-\cite{GRN3}. 
 
\section{Reductions to the two-dimensional conductivity equation} 
\label{sec:3}

In this Section we construct simple Moutard-type transforms for the conductivity equations (\ref{eq:hc1}), (\ref{eq:conj1.3}) as reductions of Moutard-type transform for the generalized analytic functions; see Section~\ref{sec:2}. 

In this Section we assume that $D$ is an open simply connected domain in $\CC\cong\RR^2$.

\begin{lemma}
\label{lem:q_sig}
A regular complex-valued function $q(z)$ admits representation (\ref{eq:hc5}) in $D$ with a positive 
$\sigma(z)$ if and only if
\begin{equation}
\label{eq:red1}
\partial_{\bar z} q(z) = \partial_{z}\overline{q(z)}, \ \ z\in D.
\end{equation}
\end{lemma}
This statement follows directly from the property that (\ref{eq:red1}) is the compatibility condition for  (\ref{eq:hc5}) and from the formula
$$
\sigma(z) =\sigma(z_0)\exp\left[-2\int\limits_{z_0}^{z} \left[ q(\zeta)d \zeta  +  \overline{q(\zeta)}d \bar\zeta  \right] \right], \ \ \mbox{where} \ \ \sigma(z_0)>0.
$$
In addition, in Lemma~\ref{lem:q_sig}, for bounded $D$, the regularity assumption (\ref{eq:assumpt1}) on  $q$ corresponds to the regularity assumptions (\ref{eq:assumpt2}), (\ref{eq:assumpt3}) on $\sigma$.

Let us define the following two special solutions $f^+_R$ and $f^+_I$ of equation~(\ref{eq:gan2}), where $q$ is given by (\ref{eq:hc5}) with a regular positive $\sigma$:
\begin{equation}
\label{eq:psiplus}
 f^+_R=\sqrt{\sigma(z)}, \ \  f^+_I=\frac{i}{\sqrt{\sigma(z)}}.
\end{equation}

\begin{lemma}
\label{lem:1}
A regular complex-valued function $\psi(z)$ satisfies equation (\ref{eq:gan1}) with $q(z)$ given by  (\ref{eq:hc5}) with a positive $\sigma(z)$  
if and only if there exists a real-valued solution $u(z)$ of (\ref{eq:hc1})  such that
\begin{equation}
\label{eq:red1.1}
\psi(z) = \sigma^{1/2}(z) \partial_{z} u(z), \ \  \overline{\psi(z)} = \sigma^{1/2}(z) \partial_{\bar z} u(z).
\end{equation}
In addition,
\beq
\label{eq:red1.2}
u = -i\omega_{\psi,f^+_I},
\eeq
where $f^+_I$ is defined in (\ref{eq:psiplus}), $\psi$ is defined in (\ref{eq:red1.1}), $\omega_{\psi,\psi+}$ is defined 
via (\ref{eq:k1}).
\end{lemma}

Note also that 
\begin{equation}
\label{eq:red1.2.5}
\psi=\frac{1}{2}\sigma^{-1/2}(I_1-i I_2), \ \ I_1=\sigma\frac{\partial u}{\partial x_1}, \ \  I_2=\sigma\frac{\partial u}{\partial x_2},
\end{equation}
where $\psi$, $u$ are the functions of (\ref{eq:red1.1}), and $I$ is the current for the conductivity equation (\ref{eq:hc1}), see formula (\ref{eq:hc1.1}).

\begin{lemma}
\label{lem:1bis}
For the conductivity equation (\ref{eq:hc1}) with regular positive $\sigma$ the following formula holds:
\beq
\label{eq:red1.3}
v = -i\omega_{\psi, f^+_R},
\eeq
where $v(z)$ is the stream function associated via (\ref{eq:conj1.2}),  (\ref{eq:conj1.3}) with real-valued $u(z)$ satisfying  (\ref{eq:hc1}), $f^+_R$ is defined in (\ref{eq:psiplus}), $\psi$ is defined in (\ref{eq:red1.1}).
\end{lemma}

Lemmas~\ref{lem:1}, ~\ref{lem:1bis} are proved in Section~\ref{sec:4}.

\begin{theorem}
\label{thm:1}
Let $q(z)$ be given by (\ref{eq:hc5})  in $D$ with a positive regular $\sigma(z)$. 
Let the transform $q\rightarrow\tilde q$, $\psi\rightarrow\tilde\psi$ be defined by:
\begin{equation}
\label{eq:moutard1}
\tilde q = \mathcal{M} q= q + \frac{f\overline{f^+}}{\omega_{f,f^+}}, \ \ 
\tilde\psi=\mathcal{M} \psi=
\psi-\frac{\omega_{_{\psi,f^+}}}{\omega_{f,f^+}}\,f,
\end{equation}
where $\psi$ denotes an arbitrary solution of  (\ref{eq:gan1}), $f$ is a fixed solution of equation (\ref{eq:gan1}), $f^+=f^+_R$ or $f^+ = f^+_I$,
where $f^+_R$ and $f^+_I$ are defined in (\ref{eq:psiplus}).

Then $\tilde\psi$ satisfies the Moutard-transformed equation (\ref{eq:gan3}), and $\tilde q$ admits the representation 
\begin{equation}
\label{eq:hc7}
\tilde q = -\frac{1}{2}\partial_{z}\log(\tilde\sigma),
\end{equation}
where
\begin{equation}
\label{eq:hc8}
\tilde\sigma = \left\{\begin{array}{ll} - \frac{\displaystyle\sigma}{\displaystyle\omega_{f,f^+_R}^2} & \mbox{if}  \ \  f^+=f^+_R,  \\ \\
- \displaystyle\sigma \displaystyle\omega_{f,f^+_I}^2 & \mbox{if} \ \ f^+=f^+_I.
\end{array}  \right.
\end{equation}
In addition, the following Moutard-transformed conjugate pair of conductivity equations holds:
\begin{equation}
\label{eq:hcm1}
\mbox{div}\big({\tilde\sigma}\nabla \tilde u \big)=0 \ \ \mbox{in} \ \  D,
\end{equation}
\begin{equation}
\label{eq:hcm1bis}
\mbox{div}\big({\tilde\sigma}^{-1}\nabla \tilde v \big)=0 \ \ \mbox{in} \ \  D,
\end{equation}
where
\begin{equation}
\label{eq:moutard2}
\tilde u = -i\omega_{\tilde\psi,\hat f^+_I}, \ \ \hat f^+_I=\frac{i}{\sqrt{\tilde\sigma}},
\end{equation}
\begin{equation}
\label{eq:moutard2bis}
\tilde v = -i \omega_{\tilde\psi,\hat f^+_R}, \ \ \hat f^+_R= \sqrt{\tilde\sigma}.
\end{equation}
\end{theorem}

The following scheme summarizes the Moutard-type transforms for the conductivity equations (\ref{eq:hc1}), (\ref{eq:conj1.3}) given in Theorem~\ref{thm:1}:

\begin{align}
&\xymatrix{ \sigma
\ar[rr]^{(\ref{eq:hc8})}_{\{f,f^+\}}   
& & 
\tilde\sigma
\ar[rr]^{(\ref{eq:hc7})}
& &  
\tilde q,}\nonumber\\
&\xymatrix{
\sigma,u  \ar[rr]^{(\ref{eq:hc5}),(\ref{eq:red1.1})}
& & 
q,\psi \ar[rr]^{(\ref{eq:moutard1})}_{\{f,f^+\}} 
& & \tilde q,\tilde\psi,}
\label{eq:pr2:1}
\\
&\xymatrix{
\tilde\sigma,\tilde\psi \ar[rr]^{(\ref{eq:moutard2}), (\ref{eq:moutard2bis} )  }
& & 
\tilde u, \tilde v.
& & 
\nonumber
}
\end{align}
The point is that each step in scheme (\ref{eq:pr2:1}) is given by quadratures. 

Theorem~\ref{thm:1} is proved in Section~\ref{sec:4}.

\begin{remark}
In Theorem~\ref{thm:1} we have the following two important cases:

\begin{enumerate}[(i)]
\item If $\omega_{f,f^+}$ has no zeroes in $D$, then $\tilde\sigma$ arising in (\ref{eq:hc8}) is a regular positive function in $D$. If $D$ is bounded, then $\omega_{f,f^+}$ can be always defined without zeroes by an appropriate choice of integration constant. 

In addition, in a similar way with the simplest case of Section~\ref{sec:2}, if $D$ is a simply connected open bounded domain with $C^1$-boundary, $\sigma$ satisfies (\ref{eq:assumpt2}),  (\ref{eq:assumpt3}), $f \in W^{1,p}(D)$, then the transformed coefficient $\tilde\sigma$ satisfies  (\ref{eq:assumpt2}),  (\ref{eq:assumpt3}).
 
\item If $\omega_{f,f^+}$ has zeroes in $D$, then $\tilde\sigma$ arising in (\ref{eq:hc8}) is non-negative and has either zeros or poles in $D$. In these singular cases the standard methods for solving the conductivity equation (\ref{eq:hcm1}) does not work; but these both singular cases are  interesting and relevant for physical problems. The point is that the Moutard-type transform  of Theorem~\ref{thm:1} generating $\tilde\sigma$ simultaneously provides a method for solving equations (\ref{eq:hcm1}), (\ref{eq:hcm1bis}).
\end{enumerate}
\end{remark}

\section{Simple Moutard transforms for the multidimensional conductivity equation}
\label{sec:5}

Let $\Mt_I$ and $\Mt_R$  denote the  Moutard-type transforms of Theorem~\ref{thm:1} for $f^+=f^+_I$ and $f^+=f^+_R$, respectively. In the next Theorem we give an explicit local realization of the transform $\Mt_I$ for conductivity equation  (\ref{eq:hc1}) and an explicit local realization of the transform $\Mt_R$ for the conjugate equation (\ref{eq:conj1.3}). However, the action of $\Mt_I$ on the solutions of (\ref{eq:conj1.3}) and   $\Mt_R$ on the solutions of (\ref{eq:hc1}) is non-local and requires one quadrature. 

\begin{theorem}
\label{thm:2}
Suppose that $f^{+}=f^{+}_I$ in Theorem~\ref{thm:1} in Section~\ref{sec:3}. Then the Moutard-type transform of Theorem~\ref{thm:1} for the two-dimensional conductivity equation (\ref{eq:hc1}) is reduced to the transform 
\begin{align}
\label{eq:mdm1}
&\sigma\rightarrow\tilde\sigma=\Mt_I\sigma=u_1^2 \sigma, \\ 
&u\rightarrow\tilde u=\Mt_I u =u_1^{-1} u,\nonumber
\end{align}
where $u_1=-i\omega_{f,f^{+}_I}$, $u(x)$ is an arbitrary real solutions of (\ref{eq:hc1}), and  $\tilde u(x)$ satisfies the Moutard transformed conductivity equation (\ref{eq:hcm1}).

Suppose that $f^{+}=f^{+}_R$ in Theorem~\ref{thm:1} in Section~\ref{sec:3}. Then the Moutard-type transform of Theorem~\ref{thm:1} for the conjugate two-dimensional conductivity equation (\ref{eq:conj1.3}) is reduced to the transform 
\begin{align}
\label{eq:mdm2}
&\sigma\rightarrow\tilde\sigma=\Mt_R\sigma=v_1^{-2} \sigma, \\ 
&v\rightarrow\tilde v=\Mt_R v =v_1^{-1} v,\nonumber
\end{align}
where $v_1=-i\omega_{f,f^{+}_R}$, $v(x)$ is an arbitrary real solution of (\ref{eq:conj1.3}), and  $\tilde v(x)$ satisfies the Moutard transformed conjugate conductivity equation (\ref{eq:hcm1bis}).
\end{theorem}

Due to formulas (\ref{eq:red1.2}), (\ref{eq:red1.3}) in Lemmas~\ref{lem:1}, \ref{lem:1bis}, functions  $u_1(x)$, $v_1(x)$ in Theorem~\ref{thm:2} are fixed real solutions of (\ref{eq:hc1}),  (\ref{eq:conj1.3}), respectively.

Note that the Moutard-type transform (\ref{eq:mdm1}) admits a direct extension to the conductivity equation 
\begin{equation}
\label{eq:mdhc1}
\mbox{div}\big({\sigma(x)}\nabla u(x) \big)=0, \ \ x=(x_1,x_2,\ldots,x_d), \ \ x\in D\subseteq\RR^d, 
\end{equation}
in dimension $d\ge 1$ (and, in particular, in dimension $d=3$). 

\begin{theorem}
\label{thm:e}
Let $\sigma$ be a real positive regular function in $D$, where $D$ is an open domain in $\RR^d$, $d\ge 1$. Let 
the transform $\sigma\rightarrow\tilde\sigma$, $u\rightarrow\tilde u$ be defined by
\begin{align}
\label{eq:mdm3}
&\tilde\sigma=\mathcal{M}\sigma=w^2 \sigma, \\ 
&\tilde u=\mathcal{M} u =w^{-1} u,\nonumber
\end{align}
where $u$ denotes an arbitrary solution of (\ref{eq:mdhc1}), and $w$ is a fixed solution of (\ref{eq:mdhc1}). Then the following Moutard-transformed 
conductivity equation holds:
\begin{equation}
\label{eq:mdhc2}
\mbox{div}\big({\tilde\sigma(x)}\nabla \tilde u(x) \big)=0, \ \  x\in D\subseteq\RR^d.
\end{equation}
\end{theorem}

Theorems~\ref{thm:2}, \ref{thm:e} are proved in Section~\ref{sec:4}.

The following result shows that subsequent application of Moutard-type transforms from Theorem~\ref{thm:e}, and, as a corollary, subsequent application of transforms from Theorem~\ref{thm:2} of the type $\Mt_I$, only, or of the type $\Mt_R$, only, does not generate new more complicated transforms. 
\begin{proposition}
\label{prop:1}
The following composition formula holds:
\begin{equation}
\label{eq:comp1}
\Mt_{\tilde\sigma,\tilde u_2} \circ \Mt_{\sigma,u_1} = \Mt_{\sigma, u_2}, \ \ \mbox{where} \ \  \tilde\sigma = \Mt_{\sigma,u_1}\sigma,\ \ \tilde u_2 = \Mt_{\sigma,u_1} u_2,
\end{equation}
where $\Mt=\Mt_{\sigma,w}$ denotes the Moutard-type transform defined by (\ref{eq:mdm3}), $u_1$, $u_2$ are arbitrary fixed solutions of (\ref{eq:mdhc1}).
\end{proposition}
In particular, from (\ref{eq:mdm3}), (\ref{eq:comp1}) it follows that 
\begin{equation}
\label{eq:comp2}
\Mt_{\tilde\sigma,\tilde u_2} \circ \Mt_{\sigma,u_1} = \mbox{id}, \ \ \mbox{if} \ \ u_2=1.
\end{equation}
Proposition~\ref{prop:1} is proved in Section~\ref{sec:4}.

On the other hand, subsequent application of transforms from  Theorem~\ref{thm:2} of the type $\Mt_I$ and then $\Mt_R$ (or $\Mt_R$ and then $\Mt_I$) yields already new transformations.

\textbf{Examples} 

1) Theorem~\ref{thm:e} implies, for example, that the multidimensional conductivity equation (\ref{eq:mdhc1}) with $\sigma=w^2$, where $w$ is a real harmonic function in $D$, is integrable in the sense that all its solutions $u$ are of the form $u = w^{-1}\phi $, where $\phi$ is an arbitrary real harmonic function in $D$. 

2) Theorem~\ref{thm:2} implies, for example, that the two-dimensional conductivity equation (\ref{eq:hc1}) in an open simply connected domain $D$ with $\sigma=w^{-2}$, where $w$ is a real harmonic function in $D$, is integrable in the sense that all its solutions $u$ are of the form
\begin{align}
&u(x)=\label{eq:trans1} \\
&=-\int\limits_{x_0}^{x}\bigg(\big[ w(\xi) \partial_{\xi_2} \phi(\xi) - \phi(\xi) \partial_{\xi_2} w(\xi) \big]d\xi_1-
\big[ w(\xi) \partial_{\xi_1} \phi(\xi) - \phi(\xi) \partial_{\xi_1} w(\xi) \big]d\xi_2\bigg)+c,
 \nonumber
\end{align}
where $\phi$ is an arbitrary real harmonic function in $D$, $x, x_0\in D$, $x_0$ is fixed, $c$ is an arbitrary real constant. Here we used formulas (\ref{eq:mdm2}) and formulas  (\ref{eq:hc1.1}), (\ref {eq:conj1.2}).

3) Theorem~\ref{thm:2} also implies, for example, that the two-dimensional conductivity equation (\ref{eq:hc1}) in an open simply connected domain $D$ with $\sigma=w^{-2}\, u_1^2$, where $w$ is a real harmonic function in $D$, $u_1$ is given by (\ref{eq:trans1}) with fixed $\phi=\phi_1$ and $c=c_1$ is integrable in the sense that all its solutions $U$ are of the form $U=u_1^{-1} \, u$, where $u$ is given by (\ref{eq:trans1}).  Here we used formulas (\ref{eq:mdm1}) with $\sigma$ and $u$ of Example~2.

For simplicity, in Examples 1-3 one can assume that $w$ and $u_1$ have no zeroes in $D$, but, formally, these examples are also valid if $w$ or $u_1$ have zeroes in $D$.
 
4) More generally, applying subsequently transforms of Theorem~~\ref{thm:2}, where applications of $\Mt_R$ are followed by applications of  $\Mt_I$ and vice verse, we obtain a very large class of integrable in quadratures two-dimensional conductivity equations in a open simply connected domain $D$.  

\section{Relations to the Schr\"odinger equation}
\label{seq:schr}

It is well-known that the substitution 
\begin{equation}
\label{eq:sch1}
u= \sigma^{-1/2} \psi,
\end{equation}
reduces the multidimensional conductivity equation (\ref{eq:mdhc1}) to the Schr\"odinger equation at zero energy:
\begin{equation}
\label{eq:sch2}
-\Delta\psi(x) + Q(x) \psi(x) =0 , \ \ x\in D\subseteq\RR^d, \ \ d\ge1,
\end{equation}
where
\begin{equation}
\label{eq:sch3}
Q(x) =  \frac{\Delta \big(\sigma^{1/2}(x) \big)}{\sigma^{1/2}(x)}.
\end{equation}
\begin{proposition}
\label{prop:2}
a) The potential $Q$ defined by (\ref{eq:sch3}) is invariant with respect to the transforms $\sigma\rightarrow\tilde\sigma $ defined in (\ref{eq:mdm3}).\\
b) In the two-dimensional case the potential  $Q$ defined by (\ref{eq:sch3}) is not invariant with respect to the transforms $\sigma\rightarrow\tilde\sigma$ defined in (\ref{eq:mdm2}). In addition, $Q$ is not invariant with respect to transform $\sigma\rightarrow \sigma^{-1}$ relating the coefficients in the conjugate pair (\ref{eq:hc1}), (\ref{eq:conj1.3}) of the two-dimensional conductivity equations.
\end{proposition}

Note that Proposition~\ref{prop:2} is in a good agreement with the fact, that no non-trivial Moutard-type transform is known for the Schr\"odinger equation (\ref{eq:sch2}) in dimension $d\ge3$, whereas the the Schr\"odinger equation (\ref{eq:sch2}) in dimension $d=2$ admits a quite rich family of Moutard transforms; see, for example, \cite{Mout}, \cite{TT}.

Item a) of Propositions~\ref{prop:2} is proved by the following calculation:
$$
\tilde Q = \frac{\Delta \big(\tilde\sigma^{1/2})}{\tilde\sigma^{1/2}} = \frac{\Delta \big( w \sigma^{1/2} \big)}{w \sigma^{1/2}}= \frac{ w \Delta \big(\sigma^{1/2} \big)}{w \sigma^{1/2}} +  
\frac{ \big( \Delta w) \sigma^{1/2} + 2 \big(\nabla w\big) \big(\nabla  \sigma^{1/2}\big)  }{w \sigma^{1/2}}=
$$
\begin{equation}
\label{eq:sch4}
=Q +\frac{\sigma \Delta w +  \big(\nabla w\big) \big(\nabla\sigma\big)}{w\sigma}=
Q +\frac{\mbox{div}\big(\sigma \nabla w \big)}{w\sigma}=Q.
\end{equation}
Here we used, in particular, that $w$ satisfies the multidimensional conductivity equation (\ref{eq:mdhc1}).

Item b) of Propositions~\ref{prop:2} follows from Examples~1 and 2 given in Section~\ref{sec:5} and the fact that $w^{-1}$ is not harmonic, in general, for harmonic $w$ (where $\tilde\sigma = w^2$ in Example~1 and $\tilde\sigma = w^{-2}$ in Example~2 with initial $\sigma\equiv1$).

Finally, consider the equations 
\begin{align}
\label{eq:ga1}
&-\mbox{div}\big({\sigma(x)}\nabla u(x) \big) + Q_1(x) u(x)=0,& \ \ &x\in D\subseteq\RR^d,&\\
\label{eq:ga11}
&-\mbox{div}\big({\sigma(x)}\nabla w(x) \big) + Q_2(x) w(x)=0,& \ \ &x\in D\subseteq\RR^d.&
\end{align}
Then (at least formally)
\begin{align}
\label{eq:ga2}
&-\mbox{div}\big({\tilde\sigma(x)}\nabla \tilde{u}(x) \big)+ q(x)\tilde u(x)=0,&  \ \ \  \ &x\in D\subseteq\RR^d,&
\end{align}
where 
\begin{equation}
\label{eq:ga3}
\tilde\sigma(x)=w^2(x) \sigma(x), \ \ \tilde{u}(x)=w^{-1}(x) u(x), \ \ q(x) = w^2(x) (Q_1(x)-Q_2(x)),
\end{equation}
$w(x)$ is a fixed solution of (\ref{eq:ga11}), $u(x)$ is an arbitrary solution of (\ref{eq:ga1}).
For particular $\sigma$, $Q_1$, $Q_2$, $w$ such a result can be found in the literature, see e.g. \cite{Van}, \cite{AM}. 

This result for $\sigma=\mbox{const}>0$, $Q_1=Q_2$, reduces the Schr\"odinger equation at zero energy (\ref{eq:ga1}) to the conductivity equation (\ref{eq:mdhc1}). In particular, this permits to construct multidimensional integrable conductivity equations from multidimensional Schr\"odinger equations integrable at zero energy. In addition, in dimension $d=2$ a very large class of Schr\"odinger equations integrable at zero energy can be constructed via classical Moutard transform (see, e.g. \cite{TT}, \cite{NTT}, \cite{NT2}). This approach for constructing integrable conductivity equations will be developed elsewhere. 

\section{Proofs of Lemmas~\ref{lem:1},  \ref{lem:1bis}, Theorems~\ref{thm:1}-\ref{thm:e} and Proposition~\ref{prop:1}}
\label{sec:4}

\textbf{Proof of Lemma~\ref{lem:1}.}
If $\sigma$ is a real-valued regular positive function, $u$ is a real-valued regular function, and
$q$, $\psi$ are defined by (\ref{eq:hc5}) and (\ref{eq:red1.1}), respectively, then it is known that $\psi$ satisfies  
(\ref{eq:gan1}) if and only if $u$ satisfies  (\ref{eq:hc1}); see Introduction. This can be also verified by a direct calculation. 

Conversely, suppose that $q$ is defined by (\ref{eq:hc5}) with a regular real-valued positive $\sigma$, and $\psi$ satisfies (\ref{eq:gan1}). Define $u$ by (\ref{eq:red1.2}). It remains to verify that (\ref{eq:red1.1}) holds. This verification uses (\ref{eq:k1}), 
(\ref{eq:psiplus}) and consists of the following:
\begin{align}
&\partial_z u = \partial_z (-i\omega_{\psi,f^+_I})=-i\psi f^+_I=-i\psi\frac{i}{\sqrt{\sigma}}=\frac{\psi}{\sqrt{\sigma}},\\
&\partial_{\bar z} u = \partial_{\bar z} (-i\omega_{\psi,f^+_I})=-i\big(-\overline{\psi\vphantom{f^+_I} }\,\overline{f^+_I}\big)
=i\overline{\psi}\frac{-i}{\sqrt{\sigma}}=\frac{\overline{\psi}}{\sqrt{\sigma}}.
\end{align}
Lemma~\ref{lem:1} is proved.

\textbf{Proof of Lemma~\ref{lem:1bis}.}
Formulas (\ref{eq:conj1.2}) defining $v$ can be rewritten as 
\begin{equation}
\label{eq:v_der}
\partial_z v = -\frac{i}{2} (I_1- i I_2), \ \ \partial_{\bar z} v = \frac{i}{2} (I_1+ i I_2).
\end{equation}
Using formulas (\ref{eq:k1}), definition $f^+_R$ in (\ref{eq:psiplus}) and formulas (\ref{eq:red1.2.5}) one can see that (\ref{eq:v_der}) is equivalent to (\ref{eq:red1.3}).

Lemma~\ref{lem:1bis} is proved.

\textbf{Proof of Theorem~\ref{thm:1}.}
The statement that $\tilde\psi$ satisfies Moutard-transformed equation (\ref{eq:gan3}) was proved in \cite{GRN1}. 

If $q$ is defined by (\ref{eq:hc7}) with $\tilde\sigma$ defined by (\ref{eq:hc8}), then:
\begin{align}
\label{eq:thm1:1}
&\tilde q =-\frac{1}{2}\partial_z\log(\tilde\sigma) = -\frac{1}{2}\partial_z\log(\sigma) + \partial_z\log(\omega_{f,f^+_R}) =\\
& = q + \frac{f f^+_R}{\omega_{f,f^+_R}} = q + \frac{f \overline{f^+_R}}{\omega_{f,f^+_R}} \ \hspace{2cm} \mbox{if} \ \ f^+=f^+_R, \nonumber
\end{align}
\begin{align}
\label{eq:thm1:2}
&\tilde q =-\frac{1}{2}\partial_z\log(\tilde\sigma) = -\frac{1}{2}\partial_z\log(\sigma) - \partial_z\log(\omega_{f,f^+_I}) =\\
& =q - \frac{f f^+_I}{\omega_{f,f^+_I}} = q + \frac{f \overline{f^+_I}}{\omega_{f,f^+_I}} \ \hspace{2cm} \mbox{if} \ \ f^+=f^+_I. \nonumber
\end{align}
Here, we used that $f^+_R$ is real-valued and $f^+_I$ is imaginary-valued. 

In fact, calculations (\ref{eq:thm1:1}), (\ref{eq:thm1:2}) prove representations (\ref{eq:hc7}),  (\ref{eq:hc8}) for $\tilde q$ 
in (\ref{eq:moutard1}).

Formulas (\ref{eq:hcm1}), (\ref{eq:moutard2}) and (\ref{eq:hcm1bis}), (\ref{eq:moutard2bis}) follow directly from the Moutard-transformed equation (\ref{eq:gan3}), the representation 
(\ref{eq:hc7}) and Lemmas~\ref{lem:1} and \ref{lem:1bis}.

This completes the proof of Theorem~\ref{thm:1} under the assumption that $\omega_{f,f^+}$ has no zeroes in $D$. 

\textbf{Proof of Theorem~\ref{thm:2}.}
The first of formulas (\ref{eq:mdm1}) follows directly from (\ref{eq:hc8}). In view of formulas (\ref{eq:hcm1}),  (\ref{eq:moutard2}), the proof of the second of formulas  (\ref{eq:mdm1}) consists of the verification that the following identity holds:
\begin{equation}
\label{eq:thm1:3}
(u_1)^{-1} u = -i \omega_{\tilde\psi,\hat f^+_I},
\end{equation}
where $\hat f^+_I$ is defined in (\ref{eq:moutard2}), $\tilde\psi$ is defined in (\ref{eq:moutard1}), where $\psi$ is defined in (\ref{eq:red1.1}), $f^+=f^+_I$, and $f$ and $u_1$ are related by (\ref{eq:red1.1}), (\ref{eq:red1.2}) with $\psi=f$, $u=u_1$. In turn, (\ref{eq:thm1:3}) can be rewritten as:
\begin{align}
u_1\sqrt{\sigma} \partial_z\big((u_1)^{-1} u  \big) = \tilde\psi,\label{eq:thm1:4}\\
u_1\sqrt{\sigma} \partial_{\bar z}\big((u_1)^{-1} u  \big) = \overline{\tilde\psi},\label{eq:thm1:5}
\end{align}
where
\begin{equation}
\label{eq:thm1:6}
\tilde\psi = \psi - \frac{u}{u_1} f,
\end{equation}
where we used (\ref{eq:red1.2}) for $\psi$, $u$ and for $\psi=f$, $u=u_1$. We have:
\begin{equation}
u_1\sqrt{\sigma} \partial_z\big((u_1)^{-1} u  \big) = \sqrt{\sigma} \partial_z u -u \sqrt{\sigma} \frac{ \partial_z u_1}{u_1}=\psi - \frac{u}{u_1} f,
\end{equation}
where we used (\ref{eq:red1.1}) for $\psi$, $u$ and for $\psi=f$, $u=u_1$. Thus, (\ref{eq:thm1:4}) holds. Formula (\ref{eq:thm1:5})  follows, for example, from reality of $u_1$, $\sqrt{\sigma}$, $u$.

This completes the proof of  (\ref{eq:mdm1}), at least if $u_1$ has no zeroes in $D$. Formulas (\ref{eq:mdm2}) are proved in similar way, at least if $v_1$ has no zeroes in $D$.

Theorem~\ref{thm:2} is proved.

\textbf{Proof of Theorem~\ref{thm:e}.}
We have that: 
\begin{equation}
\label{eq:thm3:1}
{\tilde\sigma(x)} \partial_j\tilde u(x)=w^2\sigma  \partial_j\left(\frac{u}{w}\right) = 
\sigma w\partial_ju - \sigma u \partial_jw,
\end{equation}
\begin{equation}
\label{eq:thm3:2}
\partial_j\big({\tilde\sigma(x)} \partial_j\tilde u(x)\big) = 
w \partial_j\big(\sigma \partial_ju)\big) - u \partial_j\big(\sigma \partial_jw\big), \ \ j=1,\ldots,d.
\end{equation}
Therefore,
\begin{equation}
\label{eq:thm3:2}
\mbox{div}\big({\tilde\sigma(x)} \nabla\tilde u(x)\big) = 
w\, \mbox{div}\big(\sigma \nabla u)\big) - u\, \mbox{div}\big(\sigma \nabla w\big)=0.
\end{equation}
Theorem~\ref{thm:e} is proved, at least is $w$ has no zeroes in $D$.

\textbf{Proof of Proposition~\ref{prop:1}} We have:
$$
\tilde\sigma = u_1^2 \sigma, \ \ \tilde u = (u_1)^{-1}u , \ \ \tilde u_2 = (u_1)^{-1} u_2,
$$
therefore,
$$
\Mt_{\tilde\sigma,\tilde u_2} \circ \Mt_{\sigma,u_1} \sigma = (\tilde u_2)^2 \tilde\sigma =  u_1^{-2} u_2^2 u_1^2 \sigma = u_2^2 \sigma = \Mt_{\sigma, u_2} \sigma,
$$
$$
\Mt_{\tilde\sigma,\tilde u_2} \circ \Mt_{\sigma,u_1} u = (\tilde u_2)^{-1} \tilde u=  u_1  u_2^{-1} u_1^{-1} u = u_2^{-1} u =
\Mt_{\sigma, u_2} u.
$$
Thus, Proposition~\ref{prop:1} is proved, at least if $u_1$, $u_2$ have no zeroes in $D$.

\begin{remark}
Formally, the proofs of Theorems~\ref{thm:1}-\ref{thm:e} and Proposition~\ref{prop:1} remain valid if $\omega_{f,f^+}$, $u_1$, $v_1$, $w$, $u_2$ have zeroes in $D$, but a proper analytic picture requires subsequent investigations in these cases.
\end{remark}

\textbf{Acknowledgments.} We thank Gr\'egoire Allaire for drawing our attention to the articles  \cite{Van}, \cite{AM}, which use a reduction of equation (\ref{eq:ga1}) to equation  (\ref{eq:ga2}) via  (\ref{eq:ga11}), (\ref{eq:ga3}).

\end{document}